\documentclass[aps,pre,twocolumn,groupedaddress,showpacs]{revtex4}
\usepackage{graphicx}
\newcommand{\bz}{\mathbf{z}}

\newcommand{\bX}{\mathbf{X}}
\newcommand{\bY}{\mathbf{Y}}

\newcommand{\tri}{\triangle}

\newcommand{\sep}{ \ \ \ , \ \ \ }

\newcommand{\beq}{\begin{equation}}
\newcommand{\eeq}{\end{equation}}
\newcommand{\beqn}{\begin{eqnarray}}
\newcommand{\eeqn}{\end{eqnarray}}

\newcommand{\dd}{{\rm d}}

\newcommand{\eq}{Eq.\ }
\newcommand{\eqs}{Eqs }
\newcommand{\fig}{Fig.\ }

\newcommand{\cO}{{\cal O}}

\newcommand{\cf}{c.f.~}

\begin{document}

\title{Predicting rare events in chemical reactions: application to skin cell proliferation
}
\author{Chiu Fan Lee
}
\email{cflee@pks.mpg.de}
\affiliation{Max Planck Institute for the Physics of Complex Systems,
N\"{o}thnitzer Str.~38, 01187 Dresden,
Germany
}


\date{\today}

\begin{abstract}
In a well-stirred system undergoing chemical reactions, fluctuations in the reaction propensities are approximately captured by the corresponding chemical Langevin 
equation. Within this context, 
we discuss in this work how the Kramers escape theory can be used to predict rare events in chemical reactions. As an example, we apply our approach to a recently 
proposed model on cell proliferation with relevance to skin cancer [P.~B.~Warren, Phys. Rev. E {\bf 80}, 030903 (2009)]. In particular, we provide an analytical 
explanation for the form of the exponential exponent observed in the onset rate of uncontrolled cell proliferation.
\end{abstract}
\pacs{05.40.-a, 02.50.Ey, 82.20.Kh, 87.18.Tt}

\maketitle

\section{Introduction}

Noise is ubiquitous in systems undergoing chemical reactions. In a well-stirred system, the source of noise comes from the probabilistic nature of the reactions, and can 
be analyzed  by employing the Chemical Master Equation (CME) \cite{vanKampen_B07,Gardiner_B09}. With the help of the Kramers-Moyal expansion, a  Chemical Langevin 
Equation (CLE) can be formulated to approximate the CME \cite{Kurtz_StochProcAppl78,Gillespie_JChemPhys00,Gardiner_B09}. When the number of molecules in the system is 
small, the limitations of the
approximation have been explored in \cite{Graberth_PhysicaA83,Hanggi_PRA84}. On the other hand, when the numbers of molecules of the different chemical species in the 
system are greater than certain thresholds, the CLE constitutes a reasonable approximation to the CME \cite{Gillespie_JChemPhys00}. One big advantage of the CLE is the 
well developed analytical tools available. For instance, thermally activated escape theory (see, e.g., \cite{Caroli_JStatPhys80,Hanggi_RMP90}), such as Kramers escape theory, serves as a natural 
platform for the studies of extinction rate 
of chemical species \cite{Reichenbach_PRE06,Kessler_JStatPhys07,Kamenev_PRE08,Dykman_PRL08,Schwartz_JStatPhys09}, and transition rates between two metastable states of 
the system concerned \cite{Bialek_a00}. This is the approach adopted in this work. Besides being of general interest to chemical systems, the method discussed here is 
also relevant to cellular processes. One interesting example is the recent proposal that 
 metastability in skin cell proliferation constitutes a component in the pathogenesis of cancer \cite{Warren_PRE09}. In particular, the author in \cite{Warren_PRE09} 
observed numerically that the rate for the onset of uncontrolled cell proliferation has an exponential component that scales in a specific manner with the model 
parameters. As an illustration, we shall demonstrate how the form of the exponent observed can be explained analytically within the context of Kramers escape theory.

\section{A simple example}
\begin{figure}
\caption{The potential energy landscape corresponding to the model described in \eq (\ref{ModelA}). 
There exist two fixed points: a stable one at $n_A = M$ and an unstable one at $n_A = N$. If $n_A$ gets beyond $N$, it will diverge to infinity. The phase portrait of 
the model is depicted
under the plot.
}
\label{simple}
\begin{center}
\includegraphics[scale=.4]{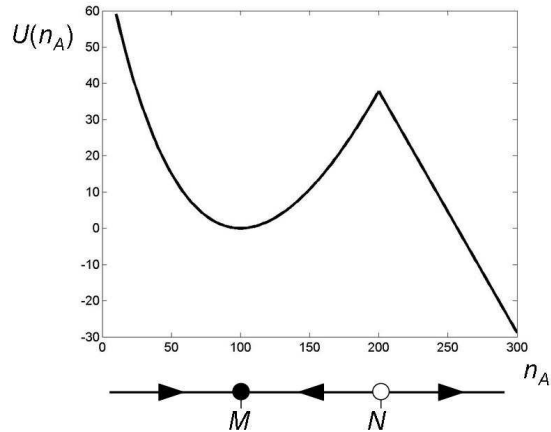}
\end{center}
\end{figure}

We will first start by considering a simple example to set up the formalism.
Consider the following set of chemical reactions:
\beq
\label{ModelA}
A  
\stackrel{\lambda}{\longrightarrow} A+A
\sep
A  +A
\stackrel{\Gamma}{\longrightarrow} A
\eeq
where $\lambda$ depends on the number of $A$ molecules, $n_a$, in the following manner:
\beq
\lambda(n_A) = 
\left\{
\begin{array}{ll}
M \Gamma \ ,\ & {\rm if \ } n_A < N
\\
2\Gamma n_A \ ,\ & {\rm otherwise} \ ,
\end{array}
\right.
\eeq
where $M$ and $\Gamma$ are constant.
In a deterministic system, the above scenario is governed by the following equation:
\beq
\dot{n}_A = \big[\lambda(n_A) -\Gamma n_A\big] n_A \ .
\eeq
In other words, if $0<n_A(t=0) < N$, then $n_A(t \rightarrow \infty) = M$. On the other hand, if $n_A(t=0) \geq N$, then $n_A(t \rightarrow \infty)$ diverges (\cf the 
phase portrait under the plot in \fig \ref{simple}).

This deterministic picture is of course incomplete due to the neglect of the intrinsic fluctuations from the reaction propensities. Such fluctuations are approximately 
captured by the following CLE \cite{Gillespie_JChemPhys00}:
\beq
\dot{n}_A = \big[\lambda(n_A) -\Gamma n_A \big] n_A +\sqrt{\lambda (n_A) n_A} w + \sqrt{\Gamma n_A^2} w' \ ,
\eeq
where $w, w'$ are Gaussian noises with zero means and unit standard deviations. Since $w$ and $w'$ are uncorrelated, we have 
\beq
\label{A1}
\dot{n}_A = \gamma f + \sqrt{2\gamma} w
\eeq
where we have introduced the following functions:
\beqn
\gamma &=& \frac{\big[\lambda(n_A) +\Gamma n_A \big] n_A}{2}
\\
f &=& \frac{2[\lambda(n_A) -\Gamma n_A]}{\lambda(n_A) +\Gamma n_A} \ .
\eeqn
Note that the form of \eq (\ref{A1}) corresponds to a Langevin equation describing a particle in a potential well under thermal perturbations in the non-inertia regime. 
Specifically, $n_A$ can be treated as the coordinate of the particle, $f$ as the  force exerted on the particle due to an underlying potential, and $\gamma$ as the 
position-dependent damping coefficient of the system.

In the region $n_A <M$, the force is
\beq
f(n_A ) = \frac{2(M- n_A)}{M+ n_A} \ .
\eeq
 The corresponding potential energy can thus be determined as
\beqn
U(n_A) &=& -\int_M^{n_A} \dd x  f(x)
\\
&=&  2(n_A-M)-4M\ln \frac{n_A+M}{2M} \ .
\eeqn
Similarly, in the region $n_A\geq M$, we have $f(n_A ) = \frac{2}{3}$ and $U(n_A) = -2n_A/3 + {\rm constant}$.
The shape of the potential energy is depicted in \fig \ref{simple}. 

We are primarily concerned with rare escape events and we thus assume that $N-M \gg 1$. 
In this scenario, if the initial state of the system is such that $n_A(t=0) = M$, then the waiting time, $\tau$, for the system to move out of the potential well, i.e., 
the waiting time for $n_A$ 
to attain the value $N$ is (\cf Sect.~7.2 in \cite{Hanggi_RMP90}):
\beqn
\tau &=& \frac{2\pi}{U'' (M)} \exp\left[2(N-M)-4M\ln \frac{N+M}{2M} \right]
\\
&=& 2\pi M\exp\left[2(N-M)-4M\ln \frac{N+M}{2M} \right] \ .
\eeqn
Note that $n_A$ will, with probability one, either go to zero or diverge \cite{Hanggi_RMP90}, and so the knowledge of the escape rate is particularly important.

\begin{figure}
\caption{(Color online) The flow lines of the cell proliferation model described in \eqs (\ref{Line1}) and (\ref{Line2}). The two fixed points of the models are depicted 
by the gray circle and square (\cf \eqs (\ref{X_def}) and (\ref{Y_def})).  The red (gray) wriggly line depicts schematically a possible escape trajectory. 
The numerical values for the model parameters are $\alpha = 2$, $\beta =10$,  $r=0.08$, $n_0=200$, $\Gamma=0.045$, $\rho_0 = 0.22$ and $\rho_1=0.26$ \cite{Warren_PRE09}.
}
\label{main3}
\begin{center}
\includegraphics[scale=.45]{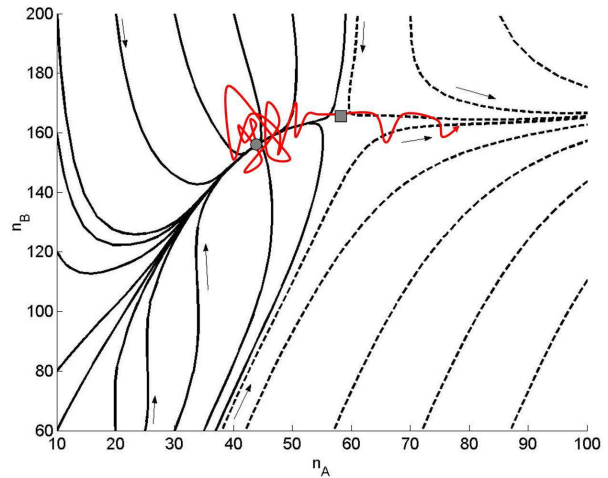}
\end{center}
\end{figure}

\section{Skin cell proliferation}
We now move onto discussing a model for skin cell proliferation.
The model we study is based on the single progenitor cell model introduced in \cite{Clayton_Nature07, Klein_PRE07}. This model was then generalized in \cite{Warren_PRE09} to account for the homeostasis of the system. Furthermore, the author in \cite{Warren_PRE09} suggests that the 
escape of the system from the homeostatic basin due to rare stochastic fluctuations plays a role in  
uncontrolled cell proliferation. This is of important relevance to the study of skin cancer. Specifically, there are two basal layer cell types in this model: progenitor 
cells $A$ and postmitotic cells 
$B$. These two types of cells proliferate according to following scheme:
\beqn
\label{Line1}
A  
\stackrel{\lambda_1}{\longrightarrow}  A+A
&,&
A  
\stackrel{\lambda_2}{\longrightarrow} A+B
\\
\label{Line2}
A  
\stackrel{\lambda_3}{\longrightarrow} B+B
& ,&
B  
\stackrel{\Gamma}{\longrightarrow} \emptyset \ .
\eeqn
The first three processes represent the different progenitor cell division pathways, and the fourth represents postmitotic cells leaving the basal layer. 
In the model 
above, $\Gamma$ is a constant and $\lambda_i$ are defined as follow:
\beqn
\lambda_1 &=& \lambda(n) r(1-q(\rho))
\\
\lambda_2 &=& \lambda(n) (1-2r)
\\
\lambda_3 &=& \lambda(n) r(1+q(\rho))
\eeqn
where
\beqn
n = n_A +n_B & , &
\rho = \frac{n_A}{n}
\\
\lambda (n) = \lambda_0 \left( \frac{n_0}{n} \right)^2
& , &
\lambda_0 = \frac{\Gamma (1-\rho_0)}{\rho_0}
\eeqn
and 
\beq
\label{q_def}
q (\rho) = \tanh \left[ \frac{10 \rho_0 (1-\rho_0)(\rho-\rho_0)(\rho_1 -\rho)}{\rho(1-\rho)(\rho_1 -\rho_0)} \right] \ .
\eeq
Note that the constants $n_0$, $\rho_0$ and $\rho_1$ represent the initial number of cells, the fraction of progenitor cells at the stable fixed point (marked by the 
gray circle in \fig \ref{main3}), and the fraction of progenitor cells at the unstable fixed point (marked by the gray square in \fig \ref{main3}), respectively. 
An experimentally motivated set of parameters for this model is shown in the caption of \fig \ref{main3}.
We refer the readers to \cite{Clayton_Nature07, Klein_PRE07, Warren_PRE09} for more detailed  physiological interpretations of the different processes. Here, we will 
only note that a divergence of the total cell density, $n_A+n_B$, signifies the onset of uncontrolled cell proliferation. In \cite{Warren_PRE09}, the author employs the 
standard Gillespie kinetic Monte Carlo algorithm \cite{Gillespie_JPhysChem77} to perform stochastic simulations of the model, and finds that the escape rate has an 
exponential component of the form $\exp (-4.6 \times n_0\tri \rho^2)$ where $\tri \rho \equiv \rho_1-\rho_0$ denotes the difference in the fractions of progenitor cells at the saddle point and at the fixed point. We will now demonstrate how the Kramers escape theory 
accounts for the exponent observed.

We shall first look at the deterministic case, where the chemical reaction scheme in \eqs (\ref{Line1}) and (\ref{Line2}) lead to the following set of ordinary 
differential equations:
\beq
\dot{n}_A = -2\lambda r q n_A
\sep
\dot{n}_B = \lambda (1+2r q)n_A -\Gamma n_B \ .
\eeq
Setting the L.H.S.~in the equations above to zero, we find two nontrivial fixed points:
\beqn
\label{X_def}
\bX &=& \big(n_0\rho_0\ ,\ (1-\rho_0)n_0 \big)
\\
\label{Y_def}
\bY &=& \left(n_0\sqrt{\frac{\lambda_0 \rho_1^3}{\Gamma (1-\rho_1)}}\ ,\ 
n_0\sqrt{\frac{\lambda_0 \rho_1 (1-\rho_1)}{\Gamma }}
 \right) \ .
\eeqn
These fixed points are denoted by a gray circle and a gray square respectively in \fig \ref{main3} along with flow lines.

The corresponding CLE for this system is \cite{Gillespie_JChemPhys00}:
\beqn
\label{ona}
\dot{n}_A &=& (\lambda_1-\lambda_3) n_A+ \sqrt{\lambda_1 n_A} w_1 -\sqrt{\lambda_3 n_A} w_3
\\
\label{onb}
\dot{n}_B &=&  (\lambda_2+2\lambda_3) n_A-\Gamma n_B
\\
&&
+ \sqrt{\lambda_2 n_A} w_2 +2\sqrt{\lambda_3 n_A} w_3 -\sqrt{\Gamma n_B} w_4 \ ,
\eeqn
where the $w_i$ are again Gaussian noises with zero means and unit standard deviations. 
As among the four independent Gaussian noise terms, only $w_3$ are common in both equations, we can thus simplify the above equation to the followings:
\beqn
\dot{n}_A &=& (\lambda_1-\lambda_3) n_A+ \sqrt{(\lambda_1 +\lambda_3)n_A} w_1 
\\
\dot{n}_B &=&  (\lambda_2+2\lambda_3) n_A-\Gamma n_B
\\
\nonumber
&&
+ \sqrt{(\lambda_2+4\lambda_3) n_A+ \Gamma n_B} \left[\sigma w_1 +\sqrt{1-\sigma^2} w_2\right]
\eeqn
where
\beq
\label{def_sigma}
\sigma = \frac{2\lambda_3 n_A}{\sqrt{n_A(\lambda_1+\lambda_3)[(\lambda_2+4\lambda_3)n_A+\Gamma n_B]}} 
\eeq
corresponds to the correlation between the two fluctuation processes.

Note that in dimensions higher than one, 
one cannot in general represent the force fields as the gradients of a potential, i.e., the force is not conservative. Although a potential energy cannot be constructed 
here, it is still possible to obtain a scalar function that serves to determine the exponent in the Arrhenius term associated to the escape process 
\cite{Matkowsky_SIAMJApplMath77}. This can be achieved by solving a second-order boundary value problem, and usually can only be done numerically. Here, we will avoid 
this numerical challenge and aim to proceed analytically by making a series of approximations to the above CLE.

\begin{figure}
\caption{The magnitude of the correlation $\sigma$ as a function of $n_A$ and $n_B$ around the two fixed points indicated again by the square and the circle. The broken 
line denotes the escape path corresponding to the analytical calculation in \eq (\ref{scaling}), and the solid line denotes the escape path obtained numerically (\cf the 
discussion before \eq (\ref{U2})), which corresponds to the result in \eq (\ref{exp2}).
}
\label{sigma}
\begin{center}
\includegraphics[scale=.45]{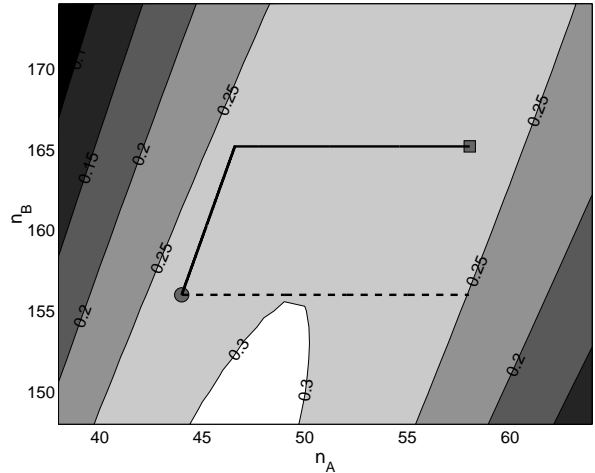}
\end{center}
\end{figure}

\begin{figure}
\caption{
The force vector fields, $(f_A,f_B)$, according to \eqs (\ref{force_fields}). The magnitudes of the vectors are scaled up uniformly for visual clarity.
}
\label{vec}
\begin{center}
\includegraphics[scale=.45]{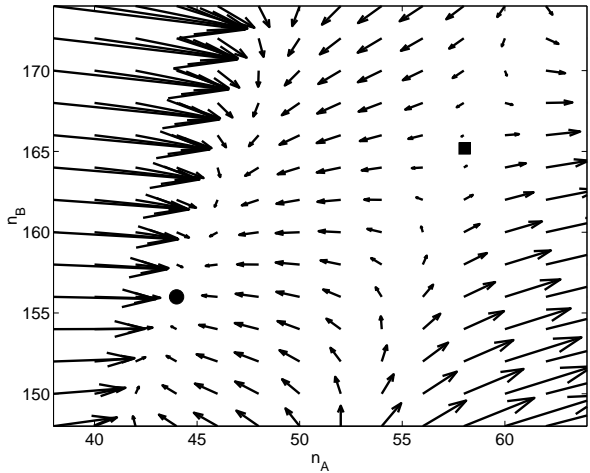}
\end{center}
\end{figure}

As aforementioned, the Gaussian noises associated to the two coordinates are correlated.
In \fig \ref{sigma}, we show the magnitude of $\sigma$ around the two fixed points, which is bounded above by 0.31. The first approximation is that we will set $\sigma$ 
to zero, i.e., we assume that the perturbations acting on $n_A$ and $n_B$ are uncorrelated. 
With this simplification, \eqs (\ref{ona}) and (\ref{onb}) can be written as:
\beq
\dot{n}_A = \gamma_A f_A +\sqrt{2 \gamma_A} w_A \sep
\dot{n}_B = \gamma_B f_B +\sqrt{2 \gamma_B} w_B\ ,
\eeq
where 
\beqn
\label{def_gamma}
\gamma_A =r \lambda n_A &,& \gamma_B = \frac{\lambda(1+2r+4rq)n_A+\Gamma n_B}{2}
\\
\label{force_fields}
f_A =  -2q  &,&  f_B = \frac{2\lambda(1+2rq)n_A-2\Gamma n_B}{\lambda(1+2r+4rq)n_A+\Gamma n_B} \ .
\eeqn
\fig \ref{vec} show the force vectors $(f_A,f_B)$ around the two fixed points, which suggests that $|f_B| \ll |f_A|$ in the region connecting $\bX$ and $\bY$. In other 
words, it is much easier for the particle to diffuse vertically than to diffuse horizontally. We will therefore ignore the second dimension and consider purely the first 
coordinate. This constitutes our second approximation and effectively collapses the problem into a one-dimensional problem. As a result, we can calculate the
corresponding potential by simply integrating over $f_A$:
\beq
U = \int 2q\left(\frac{x}{x+\hat{n}_B}\right) \dd x \ .
\eeq
where we will take $\hat{n}_B = X_B$ (\cf \eqs (\ref{q_def}) and (\ref{X_def})).
If the initial cell densities are in the metastable region around $\bX$, the rate, $R$, at which uncontrolled cell proliferation occurs will be of the form $R \propto 
\exp(-\tri U)$ \cite{Matkowsky_SIAMJApplMath77}, where
\beq
\label{z_def}
\tri U = \int_{X_A}^{Y_A} 2q\left(\frac{x}{x+\hat{n}_B}\right) \dd x \ .
\eeq
Note that our consideration effectively amounts to calculating the first passage time of a particle constrained to diffuse along the horizontal path depicted by the 
broken line in \fig \ref{sigma}.

Since throughout the range of the integration, the argument in $q$ is bounded above by 0.12, $q$ is well approximated by $\tilde{q}$ where (\cf \eq (\ref{q_def})) 
\beq
\tilde{q}(\rho) = \frac{10 \rho_0 (1-\rho_0)(\rho-\rho_0)(\rho_1 -\rho)}{\rho(1-\rho)(\rho_1 -\rho_0)}  \ .
\eeq
This simplification allows us to perform the integration in \eq (\ref{z_def}) analytically, and we find that for small $\tri \rho$,
\beqn
\label{scaling}
\tri U &=& \frac{5 \rho_0 (3-2 \rho_0)^2 }{3 (1-\rho_0) }n_0\tri \rho^2 + \cO(\tri \rho^3) 
\\
& \simeq &3.1 \times n_0 \tri \rho^2  \ ,
\eeqn
where the second approximated equality comes from substituting in the numerical values of the parameters shown in the caption of \fig \ref{main3}. 
Recall that the exponent is found numerically to be $(4.6 \times n_0 \tri \rho^2)$ \cite{Warren_PRE09}.
We have therefore recovered the scaling of the exponent with respect to $\tri \rho$. On the other hand, the prefactor we obtained is about two thirds of that observed 
from simulations. 

We will now try to incorporate the second dimension and the correlation in the fluctuations into the picture. Our strategy is to find a path that better represents the 
escape route. In the weak noise limit, such an {\it optimal} escape path encapsulates the information on the asymptotic behavior of the escape process, and in principle, can be obtained by solving a set Hamiltonian equations with the appropriate end points. 
\cite{Caroli_JStatPhys80,Maier_PRL92,Maier_PRE93,Altland_B10}. 
We find the application of the numerical procedure to the problem 
concerned challenging as the corresponding set of Hamiltonian equations are higher sensitive to the initial conditions chosen. Hence, we will instead make a crude estimate on the escape path that connects the metastable state to the saddle point  \footnote{Note that in an 
nonequilibrium system, the escape route may deviate from the saddle point \cite{Maier_PRL92,Maier_SIAMJApplMath97}. Here, we make the simple assumption that such a 
deviation is negligible.}. Specifically, as we have argued that the particle diffuses more easily along the vertical direction, we reason that as the particle goes upward in the $nA$ direction,  it should stay at the bottom of the valley with respect to the force fields along the $n_A$. We therefore start at the metastable point, $\bX$, and find the $n_A$ that minimizes 
$|f_A|$ as we move up in the $n_B$ dimension. We find that such a path corresponds to a slanted as depicted by the solid line in \fig \ref{sigma}. When the path reaches 
$Y_B$ in the $n_B$ coordinate, we simply connects it horizontally with the saddle point $\bY$ (\cf \fig \ref{sigma}). We denote this escape path by $\bz(s)$, where $0 \leq s \leq L$ corresponds to the parametrization of the curve such that $\bz(0)={\bf X}$, $\bz(L)={\bf Y}$ and $|\bz'(s)|=1$ for all $s$. Now, we collapse again the problem into one dimension by considering only fluctuation 
processes along this path. The time evolution of the particle along the path can be expressed as:
\beqn
\dot{s} &=& u\gamma_A f_A +v\gamma_B f_B
\\
\nonumber
&&+ \left[2(
u^2\gamma_A+v^2 \gamma_B+2\sigma uv\sqrt{\gamma_A \gamma_B})\right]^{1/2} w \ ,
\eeqn
where the $\gamma_{A/B}$ and $f_{A/B}$ are as expressed in \eqs (\ref{def_gamma}) and (\ref{force_fields}), $\sigma$ is defined in \eq (\ref{def_sigma}), and $(u,v)$ 
denotes the unit tangent of the curve $\bz$ at the point $(n_A,n_B)$. As a result,
the exponent in the rate describing the escape process from $s=0$ to $s=L$ is:
\beq
\label{U2}
\tri U = -\int_0^L\frac{u\gamma_A f_A +v\gamma_B f_B}{
u^2\gamma_A+v^2 \gamma_B+2\sigma uv\sqrt{\gamma_A \gamma_B}}  \ \dd s  \ .
\eeq
The numerical value is found to be
\beq
\label{exp2}
\tri U = 2.05=6.4 \times n_0 \tri \rho^2 \ ,
\eeq
which is greater than the simulation results by about 39\%. 
The discrepancy here is likely an outcome of our crude way of estimating the escape path. To improve upon this result, more sophisticated numerical approaches would be required, which is beyond the scope of this work.

\section{Conclusion}
In summary, we have discussed how the Kramers escape theory can be used to predict rare events in chemical reactions due to stochastic fluctuations. As an application, 
we have considered a model on cell proliferation and explained analytically the observed rate for the onset of uncontrolled cell growth.


\begin{thebibliography}{23}
\expandafter\ifx\csname natexlab\endcsname\relax\def\natexlab#1{#1}\fi
\expandafter\ifx\csname bibnamefont\endcsname\relax
  \def\bibnamefont#1{#1}\fi
\expandafter\ifx\csname bibfnamefont\endcsname\4relax
  \def\bibfnamefont#1{#1}\fi
\expandafter\ifx\csname citenamefont\endcsname\relax
  \def\citenamefont#1{#1}\fi
\expandafter\ifx\csname url\endcsname\relax
  \def\url#1{\texttt{#1}}\fi
\expandafter\ifx\csname urlprefix\endcsname\relax\def\urlprefix{URL }\fi
\providecommand{\bibinfo}[2]{#2}
\providecommand{\eprint}[2][]{\url{#2}}

\bibitem[{\citenamefont{van Kampen}(2007)}]{vanKampen_B07}
\bibinfo{author}{\bibfnamefont{N.~G.} \bibnamefont{van Kampen}},
  \emph{\bibinfo{title}{Stochastic Processes in Physics and Chemistry}}
  (\bibinfo{publisher}{North Holland}, \bibinfo{year}{2007}),
  \bibinfo{edition}{3rd} ed.

\bibitem[{\citenamefont{Gardiner}(2009)}]{Gardiner_B09}
\bibinfo{author}{\bibfnamefont{C.}~\bibnamefont{Gardiner}},
  \emph{\bibinfo{title}{Stochastic Methods: A Handbook for the Natural and
  Social Sciences}} (\bibinfo{publisher}{Springer}, \bibinfo{year}{2009}),
  \bibinfo{edition}{4th} ed.

\bibitem[{\citenamefont{Kurtz}(1978)}]{Kurtz_StochProcAppl78}
\bibinfo{author}{\bibfnamefont{T.}~\bibnamefont{Kurtz}},
  \bibinfo{journal}{Stochastic Processes and their Applications}
  \textbf{\bibinfo{volume}{6}}, \bibinfo{pages}{223} (\bibinfo{year}{1978}).

\bibitem[{\citenamefont{Gillespie}(2000)}]{Gillespie_JChemPhys00}
\bibinfo{author}{\bibfnamefont{D.~T.} \bibnamefont{Gillespie}},
  \bibinfo{journal}{The Journal of Chemical Physics}
  \textbf{\bibinfo{volume}{113}}, \bibinfo{pages}{297} (\bibinfo{year}{2000}).

\bibitem[{\citenamefont{Grabert et~al.}(1983)\citenamefont{Grabert, H\"{a}nggi,
  and Oppenheim}}]{Graberth_PhysicaA83}
\bibinfo{author}{\bibfnamefont{H.}~\bibnamefont{Grabert}},
  \bibinfo{author}{\bibfnamefont{P.}~\bibnamefont{H\"{a}nggi}},
  \bibnamefont{and}
  \bibinfo{author}{\bibfnamefont{I.}~\bibnamefont{Oppenheim}},
  \bibinfo{journal}{Physica A: Statistical and Theoretical Physics}
  \textbf{\bibinfo{volume}{117}}, \bibinfo{pages}{300} (\bibinfo{year}{1983}).

\bibitem[{\citenamefont{H\"{a}nggi et~al.}(1984)\citenamefont{H\"{a}nggi,
  Grabert, Talkner, and Thomas}}]{Hanggi_PRA84}
\bibinfo{author}{\bibfnamefont{P.}~\bibnamefont{H\"{a}nggi}},
  \bibinfo{author}{\bibfnamefont{H.}~\bibnamefont{Grabert}},
  \bibinfo{author}{\bibfnamefont{P.}~\bibnamefont{Talkner}}, \bibnamefont{and}
  \bibinfo{author}{\bibfnamefont{H.}~\bibnamefont{Thomas}},
  \bibinfo{journal}{Physical Review A} \textbf{\bibinfo{volume}{29}},
  \bibinfo{pages}{371} (\bibinfo{year}{1984}).

\bibitem[{\citenamefont{Caroli et~al.}(1980)\citenamefont{Caroli, Caroli,
  Roulet, and Gouyet}}]{Caroli_JStatPhys80}
\bibinfo{author}{\bibfnamefont{B.}~\bibnamefont{Caroli}},
  \bibinfo{author}{\bibfnamefont{C.}~\bibnamefont{Caroli}},
  \bibinfo{author}{\bibfnamefont{B.}~\bibnamefont{Roulet}}, \bibnamefont{and}
  \bibinfo{author}{\bibfnamefont{J.~F.} \bibnamefont{Gouyet}},
  \bibinfo{journal}{Journal of Statistical Physics}
  \textbf{\bibinfo{volume}{22}}, \bibinfo{pages}{515} (\bibinfo{year}{1980}).

\bibitem[{\citenamefont{H\"{a}nggi et~al.}(1990)\citenamefont{H\"{a}nggi,
  Talkner, and Borkovec}}]{Hanggi_RMP90}
\bibinfo{author}{\bibfnamefont{P.}~\bibnamefont{H\"{a}nggi}},
  \bibinfo{author}{\bibfnamefont{P.}~\bibnamefont{Talkner}}, \bibnamefont{and}
  \bibinfo{author}{\bibfnamefont{M.}~\bibnamefont{Borkovec}},
  \bibinfo{journal}{Reviews of Modern Physics} \textbf{\bibinfo{volume}{62}},
  \bibinfo{pages}{251} (\bibinfo{year}{1990}).

\bibitem[{\citenamefont{Reichenbach et~al.}(2006)\citenamefont{Reichenbach,
  Mobilia, and Frey}}]{Reichenbach_PRE06}
\bibinfo{author}{\bibfnamefont{T.}~\bibnamefont{Reichenbach}},
  \bibinfo{author}{\bibfnamefont{M.}~\bibnamefont{Mobilia}}, \bibnamefont{and}
  \bibinfo{author}{\bibfnamefont{E.}~\bibnamefont{Frey}},
  \bibinfo{journal}{Physical Review E} \textbf{\bibinfo{volume}{74}},
  \bibinfo{pages}{051907} (\bibinfo{year}{2006}).

\bibitem[{\citenamefont{Kessler and Shnerb}(2007)}]{Kessler_JStatPhys07}
\bibinfo{author}{\bibfnamefont{D.}~\bibnamefont{Kessler}} \bibnamefont{and}
  \bibinfo{author}{\bibfnamefont{N.}~\bibnamefont{Shnerb}},
  \bibinfo{journal}{Journal of Statistical Physics}
  \textbf{\bibinfo{volume}{127}}, \bibinfo{pages}{861} (\bibinfo{year}{2007}).

\bibitem[{\citenamefont{Kamenev and Meerson}(2008)}]{Kamenev_PRE08}
\bibinfo{author}{\bibfnamefont{A.}~\bibnamefont{Kamenev}} \bibnamefont{and}
  \bibinfo{author}{\bibfnamefont{B.}~\bibnamefont{Meerson}},
  \bibinfo{journal}{Physical Review E} \textbf{\bibinfo{volume}{77}},
  \bibinfo{pages}{061107} (\bibinfo{year}{2008}).

\bibitem[{\citenamefont{Dykman et~al.}(2008)\citenamefont{Dykman, Schwartz, and
  Landsman}}]{Dykman_PRL08}
\bibinfo{author}{\bibfnamefont{M.~I.} \bibnamefont{Dykman}},
  \bibinfo{author}{\bibfnamefont{I.~B.} \bibnamefont{Schwartz}},
  \bibnamefont{and} \bibinfo{author}{\bibfnamefont{A.~S.}
  \bibnamefont{Landsman}}, \bibinfo{journal}{Physical Review Letters}
  \textbf{\bibinfo{volume}{101}}, \bibinfo{pages}{078101}
  (\bibinfo{year}{2008}).

\bibitem[{\citenamefont{Schwartz et~al.}(2009)\citenamefont{Schwartz, Billings,
  Dykman, and Landsman}}]{Schwartz_JStatPhys09}
\bibinfo{author}{\bibfnamefont{I.~B.} \bibnamefont{Schwartz}},
  \bibinfo{author}{\bibfnamefont{L.}~\bibnamefont{Billings}},
  \bibinfo{author}{\bibfnamefont{M.}~\bibnamefont{Dykman}}, \bibnamefont{and}
  \bibinfo{author}{\bibfnamefont{A.}~\bibnamefont{Landsman}},
  \bibinfo{journal}{Journal of Statistical Mechanics: Theory and Experiment}
  \textbf{\bibinfo{volume}{2009}}, \bibinfo{pages}{P01005}
  (\bibinfo{year}{2009}).

\bibitem[{\citenamefont{Bialek}(2000)}]{Bialek_a00}
\bibinfo{author}{\bibfnamefont{W.}~\bibnamefont{Bialek}},
  \emph{\bibinfo{title}{Stability and noise in biochemical switches}}
  (\bibinfo{publisher}{MIT Press}, \bibinfo{year}{2000}), pp.
  \bibinfo{pages}{103--109}, \eprint{cond-mat/0005235}.

\bibitem[{\citenamefont{Warren}(2009)}]{Warren_PRE09}
\bibinfo{author}{\bibfnamefont{P.~B.} \bibnamefont{Warren}},
  \bibinfo{journal}{Physical Review E (Statistical, Nonlinear, and Soft Matter
  Physics)} \textbf{\bibinfo{volume}{80}}, \bibinfo{pages}{030903}
  (\bibinfo{year}{2009}).

\bibitem[{\citenamefont{Clayton et~al.}(2007)\citenamefont{Clayton, Doupe,
  Klein, Winton, Simons, and Jones}}]{Clayton_Nature07}
\bibinfo{author}{\bibfnamefont{E.}~\bibnamefont{Clayton}},
  \bibinfo{author}{\bibfnamefont{D.~P.} \bibnamefont{Doupe}},
  \bibinfo{author}{\bibfnamefont{A.~M.} \bibnamefont{Klein}},
  \bibinfo{author}{\bibfnamefont{D.~J.} \bibnamefont{Winton}},
  \bibinfo{author}{\bibfnamefont{B.~D.} \bibnamefont{Simons}},
  \bibnamefont{and} \bibinfo{author}{\bibfnamefont{P.~H.} \bibnamefont{Jones}},
  \bibinfo{journal}{Nature} \textbf{\bibinfo{volume}{446}},
  \bibinfo{pages}{185} (\bibinfo{year}{2007}).

\bibitem[{\citenamefont{Klein et~al.}(2007)\citenamefont{Klein, Doup\'{e},
  Jones, and Simons}}]{Klein_PRE07}
\bibinfo{author}{\bibfnamefont{A.~M.} \bibnamefont{Klein}},
  \bibinfo{author}{\bibfnamefont{D.~P.} \bibnamefont{Doup\'{e}}},
  \bibinfo{author}{\bibfnamefont{P.~H.} \bibnamefont{Jones}}, \bibnamefont{and}
  \bibinfo{author}{\bibfnamefont{B.~D.} \bibnamefont{Simons}},
  \bibinfo{journal}{Physical Review E (Statistical, Nonlinear, and Soft Matter
  Physics)} \textbf{\bibinfo{volume}{76}}, \bibinfo{pages}{021910}
  (\bibinfo{year}{2007}).

\bibitem[{\citenamefont{Gillespie}(1977)}]{Gillespie_JPhysChem77}
\bibinfo{author}{\bibfnamefont{D.~T.} \bibnamefont{Gillespie}},
  \bibinfo{journal}{The Journal of Physical Chemistry}
  \textbf{\bibinfo{volume}{81}}, \bibinfo{pages}{2340} (\bibinfo{year}{1977}).

\bibitem[{\citenamefont{Matkowsky and
  Schuss}(1977)}]{Matkowsky_SIAMJApplMath77}
\bibinfo{author}{\bibfnamefont{B.~J.} \bibnamefont{Matkowsky}}
  \bibnamefont{and} \bibinfo{author}{\bibfnamefont{Z.}~\bibnamefont{Schuss}},
  \bibinfo{journal}{SIAM Journal on Applied Mathematics}
  \textbf{\bibinfo{volume}{33}}, \bibinfo{pages}{365} (\bibinfo{year}{1977}).

\bibitem[{\citenamefont{Maier and Stein}(1992)}]{Maier_PRL92}
\bibinfo{author}{\bibfnamefont{R.~S.} \bibnamefont{Maier}} \bibnamefont{and}
  \bibinfo{author}{\bibfnamefont{D.~L.} \bibnamefont{Stein}},
  \bibinfo{journal}{Physical Review Letters} \textbf{\bibinfo{volume}{69}},
  \bibinfo{pages}{3691} (\bibinfo{year}{1992}).

\bibitem[{\citenamefont{Maier and Stein}(1993)}]{Maier_PRE93}
\bibinfo{author}{\bibfnamefont{R.~S.} \bibnamefont{Maier}} \bibnamefont{and}
  \bibinfo{author}{\bibfnamefont{D.~L.} \bibnamefont{Stein}},
  \bibinfo{journal}{Physical Review E} \textbf{\bibinfo{volume}{48}},
  \bibinfo{pages}{931} (\bibinfo{year}{1993}).

\bibitem[{\citenamefont{Altland and Simons}(2010)}]{Altland_B10}
\bibinfo{author}{\bibfnamefont{A.}~\bibnamefont{Altland}} \bibnamefont{and}
  \bibinfo{author}{\bibfnamefont{B.}~\bibnamefont{Simons}},
  \emph{\bibinfo{title}{Condensed Matter Field Theory}}
  (\bibinfo{publisher}{Cambridge University Press}, \bibinfo{year}{2010}),
  \bibinfo{edition}{2nd} ed.

\bibitem[{\citenamefont{Maier and Stein}(1997)}]{Maier_SIAMJApplMath97}
\bibinfo{author}{\bibfnamefont{R.~S.} \bibnamefont{Maier}} \bibnamefont{and}
  \bibinfo{author}{\bibfnamefont{D.~L.} \bibnamefont{Stein}},
  \bibinfo{journal}{SIAM Journal on Applied Mathematics}
  \textbf{\bibinfo{volume}{57}}, \bibinfo{pages}{752} (\bibinfo{year}{1997}).

\end{thebibliography}
\end{document}